# Tunable hyperbolic metamaterial cavity towards few exciton strong coupling

**LINGLING FAN,**[1] **WENYANG WU,**[1] **WENBO ZANG,**[1] **ZHUO CHEN,** [1,2,*] **AND ZHENLIN WANG**[1,2]

[1]*School of Physics and National Laboratory of Solid State Microstructures, Nanjing University, Nanjing 210093, China*
[2]*Collaborative Innovation Center of Advanced Microstructures, Nanjing University, Nanjing 210093, China*
*\*zchen@nju.edu.cn*

**Abstract:** We theoretically investigate coupling interaction between molecule excitons and whispering gallery modes (WGMs) that approaches the limit of single exciton strong coupling in hyperbolic metamaterial (HMM) cavity, composed of a dielectric core wrapped by several stacks of alternative layers of metal and dielectric. We demonstrate that associated with the excitation of the whispering gallery mode, the electric fields with resonance wavelengths that are much larger than the cavity size can be highly localized within a dielectric layer, leading to extremely small mode volumes. By using temporal coupled mode theory to model the interaction in the proposed WGMs-exciton system, we are able to demonstrate that the coupling between molecular excitons and hyperbolic cavity WGMs can reach the strong coupling regime. Furthermore, we also prove that changing both the thickness of inner core and outermost layer can lead to invariant resonance frequency while the threshold number of coupled excitons to fulfill the criteria for strong coupling remains ~ 4 in various sets of tunable HMM cavities.

## References and links

1. R. Chikkaraddy, B. de Nijs, F. Benz, S. J. Barrow, O. A. Scherman, E. Rosta, A. Demetriadou, P. Fox, O. Hess, and J. J. Baumberg, "Single-molecule strong coupling at room temperature in plasmonic nanocavities," Nature 535, 127-130 (2016)..
2. D. M. Coles, Y. Yang, Y. Wang, R. T. Grant, R. A. Taylor, S. K. Saikin, A. Aspuru-Guzik, D. G. Lidzey, J. K. Tang, and J. M. Smith, "Strong coupling between chlorosomes of photosynthetic bacteria and a confined optical cavity mode," Nat Commun 5, 5561 (2014)..
3. J. D. Plumhof, T. Stöferle, L. Mai, U. Scherf, and R. F. Mahrt, "Room-temperature Bose–Einstein condensation of cavity exciton–polaritons in a polymer," Nat Mater 13, 247-252 (2014).
4. D. Yelin, D. Oron, S. Thiberge, E. Moses, and Y. Silberberg, "Multiphoton plasmon-resonance microscopy," Opt. Express **11**(12), 1385–1391 (2003).
5. J. Z. Bernád, and G. Alber, "Photon-assisted entanglement creation by minimum-error generalized quantum measurements in the strong-coupling regime," Physical Review A 87, 012311 (2013).
6. C. Coulson, G. Christmann, P. Cristofolini, C. Grossmann, J. J. Baumberg, S. I. Tsintzos, G. Konstantinidis, Z. Hatzopoulos, and P. G. Savvidis, "Electrically controlled strong coupling and polariton bistability in double quantum wells," Physical Review B 87, 045311 (2013).
7. R. Konoike, H. Nakagawa, M. Nakadai, T. Asano, Y. Tanaka, and S. Noda, "On-demand transfer of trapped photons on a chip," Science Advances 2 (2016).
8. Y. Sato, Y. Tanaka, J. Upham, Y. Takahashi, T. Asano, and S. Noda, "Strong coupling between distant photonic nanocavities and its dynamic control," Nature Photonics 6, 56-61 (2011).
9. F. Neubrech, A. Pucci, T. W. Cornelius, S. Karim, A. Garcia-Etxarri, and J. Aizpurua, "Resonant plasmonic and vibrational coupling in a tailored nanoantenna for infrared detection," Phys Rev Lett 101, 157403 (2008).
10. L. V. Brown, K. Zhao, N. King, H. Sobhani, P. Nordlander, and N. J. Halas, "Surface-enhanced infrared absorption using individual cross antennas tailored to chemical moieties," J Am Chem Soc 135, 3688-3695 (2013).
11. D. Nau, A. Seidel, R. B. Orzekowsky, S. H. Lee, S. Deb, and H. Giessen, "Hydrogen sensor based on metallic photonic crystal slabs," Opt. Lett. 35, 3150-3152 (2010).
12. V. E. Ferry, L. A. Sweatlock, D. Pacifici, and H. A. Atwater, "Plasmonic nanostructure design for efficient light coupling into solar cells," Nano Letters 8, 4391 (2008).

13. A. E. Schlather, N. Large, A. S. Urban, P. Nordlander, and N. J. Halas, "Near-field mediated plexcitonic coupling and giant Rabi splitting in individual metallic dimers," Nano Lett 13, 3281-3286 (2013).
14. W. Wu, M. Wan, P. Gu, Z. Chen, and Z. Wang, "Strong coupling between few molecular excitons and Fano-like cavity plasmon in two-layered dielectric-metal core-shell resonators," Opt Express 25, 1495-1504 (2017).
15. P. Torma, and W. L. Barnes, "Strong coupling between surface plasmon polaritons and emitters: a review," Rep Prog Phys 78, 013901 (2015).
16. G. Khitrova, H. M. Gibbs, M. Kira, S. W. Koch, and A. Scherer, "Vacuum Rabi splitting in semiconductors," Nat Phys 2, 81-90 (2006).
17. K. J. Vahala, "Optical microcavities," Nature 424, 839-846 (2003).
18. S. Anand, M. Eryürek, Y. Karadag, A. Erten, A. Serpengüzel, A. Jonáš, and A. Kiraz, "Observation of whispering gallery modes in elastic light scattering from microdroplets optically trapped in a microfluidic channel," J. Opt. Soc. Am. B 33, 1349-1354 (2016).
19. B. Gayral, J. M. Gérard, A. Lemaître, C. Dupuis, L. Manin, and J. L. Pelouard, "High-Q wet-etched gaas microdisks containing inas quantum boxes," Appl. Phys. Lett. 75(13), 1908 (1999).
20. E. Peter, P. Senellart, D. Martrou, A. Lemaître, J. Hours, J. M. Gérard, and J. Bloch, "Exciton-Photon Strong-Coupling Regime for a Single Quantum Dot Embedded in a Microcavity," Physical Review Letters 95, 067401 (2005).
21. D. Englund, A. Majumdar, A. Faraon, M. Toishi, N. Stoltz, P. Petroff, and J. Vuckovic, "Resonant excitation of a quantum dot strongly coupled to a photonic crystal nanocavity," Phys Rev Lett 104, 073904 (2010).
22. P. Androvitsaneas, A. B. Young, C. Schneider, S. Maier, M. Kamp, S. Höfling, S. Knauer, E. Harbord, C. Y. Hu, J. G. Rarity, and R. Oulton, "Charged quantum dot micropillar system for deterministic light-matter interactions," Physical Review B 93, 241409 (2016).
23. Y. Yang, O. D. Miller, T. Christensen, J. D. Joannopoulos, and M. Soljačić, "Low-Loss Plasmonic Dielectric Nanoresonators," Nano Letters 17, 3238-3245 (2017).
24. X. Chen, Y.-H. Chen, J. Qin, D. Zhao, B. Ding, R. J. Blaikie, and M. Qiu, "Mode Modification of Plasmonic Gap Resonances Induced by Strong Coupling with Molecular Excitons," Nano Letters 17, 3246-3251 (2017).
25. P. Oskar, S. Kartik, D. O. B. John, S. Axel, and P. D. Dapkus, "Tailoring of the resonant mode properties of optical nanocavities in two-dimensional photonic crystal slab waveguides," Journal of Optics A: Pure and Applied Optics 3, S161 (2001).
26. J. S. Foresi, P. R. Villeneuve, J. Ferrera, E. R. Thoen, G. Steinmeyer, S. Fan, J. D. Joannopoulos, L. C. Kimerling, H. I. Smith, and E. P. Ippen, "Photonic-bandgap microcavities in optical waveguides," Nature 390, 143-145 (1997).
27. R. Coccioli, M. Boroditsky, K. W. Kim, Y. Rahmat-Samii, and E. Yablonovitch, "Smallest possible electromagnetic mode volume in a dielectric cavity," IEE Proceedings - Optoelectronics 145, 391-397 (1998).
28. A. Faraon, P. E. Barclay, C. Santori, K.-M. C. Fu, and R. G. Beausoleil, "Resonant enhancement of the zero-phonon emission from a colour centre in a diamond cavity," Nat Photon 5, 301-305 (2011).
29. J. P. Reithmaier, G. Sek, A. Loffler, C. Hofmann, S. Kuhn, S. Reitzenstein, L. V. Keldysh, V. D. Kulakovskii, T. L. Reinecke, and A. Forchel, "Strong coupling in a single quantum dot-semiconductor microcavity system," Nature 432, 197-200 (2004).
30. D. Wang, H. Kelkar, D. Martin-Cano, T. Utikal, S. Götzinger, and V. Sandoghdar, "Coherent Coupling of a Single Molecule to a Scanning Fabry-Perot Microcavity," Physical Review X 7, 021014 (2017).
31. P. B. Johnson, and R. W. Christy, "Optical Constants of the Noble Metals," Physical Review B 6, 4370-4379 (1972).
32. P. Gu, M. Wan, Q. Shen, X. He, Z. Chen, P. Zhan, and Z. Wang, "Experimental observation of sharp cavity plasmon resonances in dielectric-metal core-shell resonators," Applied Physics Letters 107, 141908 (2015).
33. C. F. Bohren, and D. R. Huffman, "Absorption and Scattering of Light by Small Particles," Journal of Colloid & Interface Science 98, 290-291 (1998).
34. C. Wu, A. Salandrino, X. Ni, and X. Zhang, "Electrodynamical Light Trapping Using Whispering-Gallery Resonances in Hyperbolic Cavities," Physical Review X 4 (2014).
35. S. T. Smiley, M. Reers, C. Mottolahartshorn, M. Lin, A. Chen, T. W. Smith, G. D. Steele, and L. B. Chen, "Intracellular heterogeneity in mitochondrial membrane potentials revealed by a J-aggregate-forming lipophilic cation JC-1," Proceedings of the National Academy of Sciences of the United States of America 88, 3671 (1991).
36. E. T. Jaynes, and F. W. Cummings, "Comparison of quantum and semiclassical radiation theories with application to the beam maser," Proceedings of the IEEE 51, 89-109 (2005).
37. S. Fan, W. Suh, and J. D. Joannopoulos, "Temporal coupled-mode theory for the Fano resonance in optical resonators," J. Opt. Soc. Am. A 20, 569-572 (2003).
38. F. Vaianella, and B. Maes, "Fano resonance engineering in slanted cavities with hyperbolic metamaterials," Physical Review B 94, 125442 (2016).

## 1. Introduction

One of the building blocks for quantum optics is the hybrid state of visible light and individual emitter[1] with significant cavity quantum electrodynamics effects. Profound phenomenon will emerge when the light and matter strongly couple with each other as part light, part matter, serving as the versatile platform to study fundamental science, e.g. complex natural photosynthesis process[2], Bose-Einstein condensation of polariton[3], photon entanglement[4] and bi-stability[5]. On the other hand, strong coupling will facilitate the sensing and modification of local optical field (e.g. low-loss switches and laser[6, 7], pinpoint molecule detection[8-10]) and harvest the light energy with increased coherent energy transfer rate (e.g. enhanced solar cell efficiency[11]). To fulfill the criterion of light-matter strong coupling, the rate of light transfer ($g$) between cavity and emitter must exceed the emitter scattering rate ($\gamma$) and cavity loss rate ($\kappa$), namely $2g > \kappa, \gamma$[12-15], which is challenged by hundreds-fold difference of spatial scales between visible light and individual emitter. There are mainly two routes to address the challenge. One is to lower $\kappa$ with high quality factor (Q) cavities[16]. Progress has been made in this direction, with the invention of whispering gallery spheres[17], micro-disks[18, 19], photonic crystals[20], micro-pillars[21] and nanoparticle-on-mirror geometry[22, 23]. However, this approach is limited by the size of the cavity, as g is proportional to $\frac{1}{\sqrt{V}}$ and the mode volume can be hardly below $V_\lambda = \left(\frac{\lambda}{n}\right)^3$ [24-26], putting a physical limit for promoting $g$ in this way. Another way is to lower $\gamma$ with special cryogenic systems like diamond vacancy[27] and quantum dots in semiconductor cavity[28]. Likewise, this is constrained by the complex fabrication and extreme temperature condition that are not favorable for practical application. Up to now, the single or few molecule exciton strong coupling is only realized in rarely special experimental designs[1, 29], while a synthetic study with comprehensive theory that guide to tune the structure still lacks, which is imperative.

In this paper, we demonstrate spherical hyperbolic metamaterial (HMM) cavities, composed of a dielectric core wrapped by five alternate metal/dielectric pairs, which display dipolar whispering gallery modes (WGMs), can serve as a tunable platform for strong coupling with few molecular exciton. Associated with the excitations of the dipolar WGMs, the electric fields with resonance wavelengths much larger than the cavity size can be highly localized within a dielectric layer, leading to very small mode volumes. By introducing J-aggregate into the layer of localization, we prove that the WGMs can couple to the molecular excitons, which displays Rabi splitting of 93 meV in the extinction and absorption spectrum. By utilizing the two-oscillator temporal coupled theory to describe the proposed WGMs-exciton system, we are able to confirm that such a coupling process can reach the strong coupling regime. Furthermore, we demonstrate the resonance frequency and threshold of molecule exciton are both invariant by changing the inner core and outermost layer thickness, demonstrating the tunable feature of HMM for strong coupling with few (~4) molecule exciton.

## 2. Results and discussions

Figure 1(a) inset shows the schematic of a spherical HMM cavity, consisting of a dielectric core with a dielectric core (radius: $r$ =3.8 nm) alternately wrapped by metal (thickness: $t$ = 5 nm) and dielectric layers (thickness: $d$ = 5 nm), thus the filling factor for thickness of metal layer over the multilayer unit in this structure is f = $\frac{t}{t+d}$ = 0.5. Here the dielectric core and shell both have the refractive index of $n$ = 1.25 (e.g. aerogel). The metal is silver with refraction index from experimental data[30]. Such a structure has been realized experimentally by two-step fabrication that is feasible for almost all deposition of metal/dielectric multilayer[31]. Throughout the paper, the scattering, extinction and absorption problem of sphere is analytically solved by Mie theory[32]. Using effective medium description, the permittivity tensor of metamaterial spherical cavity is represented by $\varepsilon_{eff} = \varepsilon_r \hat{r}\hat{r} + \varepsilon_t \hat{\theta}\hat{\theta} + \varepsilon_t \hat{\varphi}\hat{\varphi}$, where the relative permittivity along the radial direction $\varepsilon_r$ and the sphere surface $\varepsilon_t$ to be $\varepsilon_r = \frac{\varepsilon_m \varepsilon_d}{[\varepsilon_m(1-f)+\varepsilon_d f]} > 0, \varepsilon_t = \varepsilon_m f + \varepsilon_d(1-f) < 0$ [33] respectively, with $f$ being the filling ratio of metal layer of the multilayer period and $\varepsilon_m, \varepsilon_d$ being the permittivity of metal and dielectric, respectively. The indefinite property of the permittivity tensor enables the spherical HMM to support WGM by Wu et al.[34]. Figure 1(a) shows the extinction spectra displays a clear peak around energy E = 1.79 eV. Via Mie solution, the spectra can be decomposed into each multipolar contribution, represented by electric $a_l$ and magnetic $b_l$ ($b_l \ll a_l$, in our case) coefficients by Mie expansion. Whispering gallery mode denoted by $WGM_{l,m}$ with the same angular momentum (*l*) may have different mode orders *m*, where *m* corresponds to $m_{th}$ longest resonance wavelength and localized maxima field within the $m_{th}$ layer from the outmost shell. From Figure 1(a), it is clear that the first electric term $a_1$ dominates the extinction spectra in the wavelength range shown in the figure. To locate the resonance peak in the field contribution, we calculate the spatial distribution of the electric field enhancement ($|E/E_0|$) at resonances $WGM_{1,4}$ and $WGM_{2,2}$ respectively, which proves that the electric fields are tightly localized within the 4$^{th}$ and 2$^{nd}$ dielectric layer from outermost shell respectively, while in the former the field intensity is enhanced larger than in the later. Quality factor $Q = \frac{\lambda_{res}}{\Gamma}$, with $\Gamma$ from Fano lineshape fitting bandwidth, is found to be Q ~ 46.5 for $WGM_{1,4}$. On resonance, the mode volumes of the resonance $WGM_{1,4}$ is $8.9 \times 10^{-5} \left(\frac{\lambda}{n}\right)^3$ from $V_m = \frac{\iiint W(r) d^3 r}{\max\{W(r)\}}$, with W(r) = $\frac{1}{2}\left[Re\left[\frac{d(\varepsilon(\vec{r}))}{d\omega}\right]|E(\vec{r})|^2 + \mu|H(\vec{r})|^2\right]$ [37] being the energy density at position $\vec{r}$. Such an extremely small mode volume is due to the indefinite property of permittivity tensor, so the Purcell factor $F_p = \frac{3}{4\pi^2}\left(\frac{\lambda}{n}\right)^3 \frac{Q}{V_m}$ is about $3.9 \times 10^4$, which makes the coupling with dye molecule exciton easier.

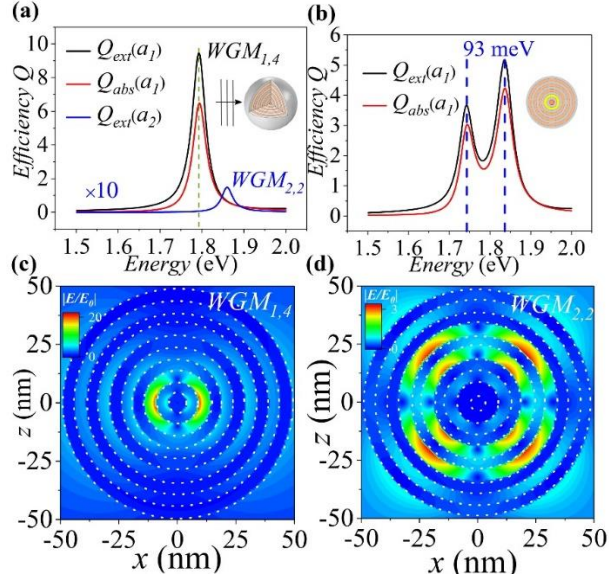

Fig. 1. (a) Extinction efficiency ($Q_{ext}$) and absorption ($Q_{abs}$) spectrum of the HMM cavity, which is decomposed into first two electric terms ($a_1, a_2$) of Mie expansion. For clarity, the contribution from $a_2$ is magnified 10 times. Inset: Schematic of a spherical HMM cavity composed of a dielectric core (radius: r = 3.8 nm) and five alternating layers of silver (thickness: t = 5 nm) and dielectric (thickness: d = 5 nm). (b) Rabi splitting with 93 meV in the extinction and absorption spectrum when J-aggregate is introduced into the field resonance layer. Inset schematically indicates that molecule exciton is put on the fourth layer from outermost shell (yellow region). (c) and (d) show the field intensity enhancement profile of $WGM_{1,4}$ and $WGM_{2,2}$ in k-E plane, respectively. Dashed lines indicate the interfaces.

So far, the HMM cavity that displays the mode of $WGM_{1,4}$ with resonance frequency at 1.79 eV, can couple with the J-aggregate molecule exciton mode. The J-aggregate[34] with central resonance frequency $\omega_e = 1.79 \, eV$ and damping rate $\gamma_e = 52$ meV is described by Lorentz oscillator model as $\varepsilon_J(\omega) = \varepsilon_\infty + \frac{f\omega_e^2}{\omega_e^2 - \omega^2 - i\gamma_e\omega}$. $\varepsilon_\infty = 1.25^2$ is the dielectric constant of background dielectric layer, $f = 0.03$ is reduced oscillator strength of excitons related with the molecule exciton number. As is shown by the inset of figure 1(b), the J-aggregate exciton is put on the fourth layer from the outermost shell and molecule exciton mode couples with the $WGM_{1,4}$ within the cavity with particular parameter $h_1 = 3.8\ nm, h_2 = 5\ nm$, which results in Rabi splitting of $\Omega$ =93 meV as is shown in extinction and absorption spectrum of molecule-doped cavity in figure 1(b). The simultaneous splitting in extinction and absorption spectrum verifies the occurrence of strong interaction between molecule exciton and HMM cavity. In figure 2(a), HMM cavity resonance frequency displays red-shift and can cross $\omega = 1.79$ eV (horizontal dashed line) when the radius of inner core $h_1$ reaches 3.8 $nm$ in the process of $h_1$ varying from 1 nm to 7 nm while the thickness of outermost layer $h_2$ staying constant. In contrast, after J-aggregate molecule exciton is introduced into the cavity, there is a distinct anti-crossing behaviour in extinction spectra from figure 2 (b) and absorption spectra from (c) when the structure is detuned by varying $h_1$. This behavior is further verified by blue-shift and anti-crossing figure 2 (d)-(f) when detuning the resonance by varying $h_2$. In both cases, we can get the distance of two anti-crossing branches at resonance equal to 93 meV, consistent with the Rabi-splitting result. This can be explained by Jaynes-Cummings picture[35]. The dispersion in anti-crossing is illustrated by $\omega_\pm = \frac{1}{2}(\omega_0 + \omega_e) \pm \frac{1}{2}\sqrt{\Omega^2 + \delta^2}$. Here $\omega_\pm$ represents the upper and lower branches, $\omega_0$ is the hyperbolic resonance frequency, which is changed when varying the structure. At resonance, $\delta = \omega_0 - \omega_e = 0$, so the distance between two peaks equal to $\Omega$, the splitting energy.

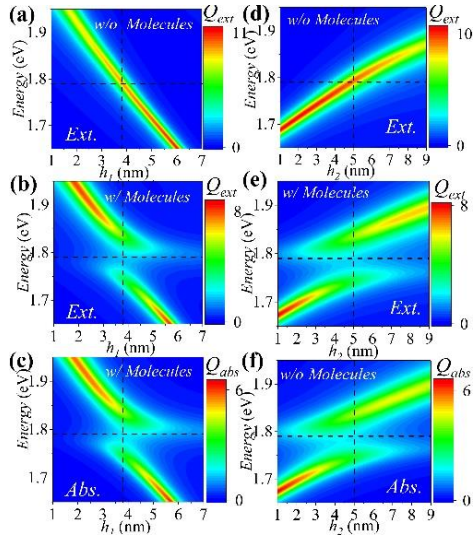

Fig. 2. Extinction spectra of the HMM cavity without (top panel (a), (d)) and with molecules (middle panel (b), (e)), and absorption spectra of the HMM cavity with molecules (bottom panel (c), (f)). There is anti-crossing at the transition energy of molecule exciton (horizontal dashed line) when molecule is put into the cavity in the yellow region of figure 1(b) inset. On the left, (a)-(c) represent the cavity structure changes $h_1$ and reaches on resonance condition when $h_1$ equals to 3.8 $nm$ (vertical dashed line), while the outer layer thickness $h_2$ is fixed. On the right, (d)-(f) represent the cavity structure changes $h_2$ and reaches resonance condition when $h_2 = 5$ nm (vertical dashed line), while the core radius $h_1$ is fixed.

To further extract the molecule exciton number involved in coupling interaction, we utilize the two-oscillator temporal coupled mode theory [13, 36] to analyze the interaction between molecule exciton and hyperbolic metamaterial cavity. As discussed above through spectrum and field profile analysis, we have known the coupling process relates hyperbolic cavity WGM mode and molecule exciton mode, which can be viewed as one bright oscillator and one dark massive oscillator in temporal coupled model, respectively. Firstly we have several assumptions about intuitive physical interaction: There is no interaction between molecules due to dilute molecule among the dielectric layer; the molecule exciton mode only couples to the cavity; the hyperbolic mode itself is no loss; only one port for incident and outgoing poynting energy flux, which means the cavity mode only couples to incident energy flow. Under such assumptions, the one-port two-oscillator temporal coupled model under these assumptions is,

$$\frac{dP_1}{dt} = -j\omega_1 P_1 + j\kappa_{1e} P_e + j\sqrt{\gamma_{1s}} S_+$$

$$\frac{dP_e}{dt} = -j\omega_e P_e - \frac{\gamma_e}{2} P_e + j\kappa_{1e} P_1$$

$$S_- = -S_+ + \sqrt{\gamma_{1s}} P_1$$

$$Q_{ext} = |S_+|^2 - |S_-|^2$$

Here $P_1, P_e$ respectively represent the HMM cavity and molecule mode amplitudes. $\omega_1, \omega_e$ are resonant frequency in their respective modes, $\gamma_e$ is the damping rate of molecular oscillator, as we have known $\omega_e$=1.79 eV, $\gamma_e$ =52 meV. In the coupling interaction between these two oscillators, $\kappa_{1e}$ represents the coupling strength between WGM of HMM cavity and molecule exciton mode. $\gamma_{1s}$ is the coupling coefficient between cavity mode and incident energy flow. $S_+$ and $S_-$ are the incident and outgoing energy flux respectively. Since extinction is the sum of scattering and absorption, equal to the amount of energy loss of incident relative to outgoing flux, extinction is proportional to the difference of $S_+$ and $S_-$, therefore extinction efficiency $Q_{ext}$ can be written as $Q_{ext} = |S_+|^2 - |S_-|^2$.

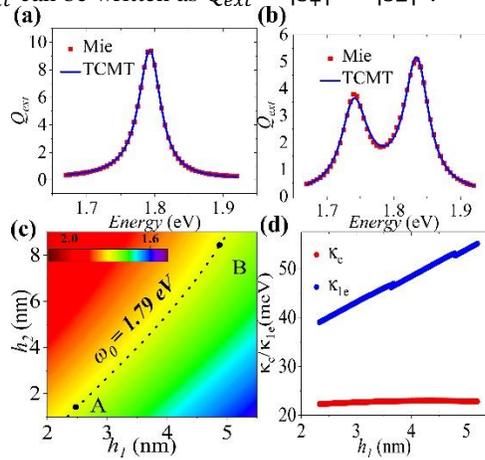

Fig. 3. (a) Extinction spectrums of WGM$_{1,4}$ in HMM cavity without molecule exciton from Mie theory (red squares) and fitting by two-oscillator temporal coupled mode theory (blue solid line). (b) Extinction spectrum of plexciton mode in the molecule-doped HMM cavity from Mie theory (blue circles) and further fitted by temporal coupled mode theory using same parameter from (a) except for $\kappa_{1e} \neq 0$. Both (a) and (b) show consistence between TCMT and Mie theory. (c) The equi-energy plot of HMM cavity resonance as a function of inner core radius $h_1$ and outer layer

thickness $h_2$. The dashed line in (c) indicates WGM with resonance of 1.79 eV contours, where at point A the mode volume is $8.99 \times 10^{-5} \left(\frac{\lambda}{n}\right)^3$ and Purcell factor is $4.27 \times 10^4$ and at point B the mode volume is $8.62 \times 10^{-5} \left(\frac{\lambda}{n}\right)^3$ and Purcell factor is $4.15 \times 10^4$. (d) The coupling strength $\kappa_{1e}$ (blue line) and $\kappa_c$ (red line) as a function of inner core radius $h_1$ for each $(h_1, h_2)$ combination on the dashed line of (c). All the combinations of $(h_1, h_2)$ fulfills strong coupling criteria $\kappa_{1e}/\kappa_c > 1$.

We use two-step fitting to get the interaction coefficients as described above. If $\kappa_{1e}$ is zero, then there is no interaction between cavity and molecule, which means the extinction is from the single HMM cavity, as is verified in figure 3(a) that the fitting extinction spectra (blue line) from figure 1 is in excellent agreement with the spectra obtained from Mie theory (red square). Using the fitting coefficients except for $\kappa_{1e}$, there is only one freedom of optimization, which is much easier and more convenient to check our assumptions. As in the figure 3(b), we get consistent fitting in the second step using the same coefficients from the first step except that $\kappa_{1e} \neq 0$. Therefore, our assumptions made above are valid in understanding the interaction between molecule exciton and hyperbolic cavity mode. Based on the optimal coefficients from TCMT fitting, the coupling strength between two oscillators in the particular structure is 47.3 meV. The system can be treated as light-matter strong coupling only when the coupling strength is larger than critical strength $\kappa_c = \sqrt{\frac{\Gamma^2 + \gamma_e^2}{8}}$, where $\Gamma$ is fano fitting loss and $\gamma_e = 52$ meV is molecule exciton damping rate. From Fano fiting of the lineshape in figure 1, $\Gamma = 38.5$ meV. So critical strength in this case is $\kappa_c = 22.8$ meV, which is smaller than the coupling strength in the coupling of two oscillators, demonstrating that the system with HMM cavity and molecule exciton has entered strong coupling regime[14].

Now we broaden our investigation for TCMT model in a fixed structure to a wide range of tunable structures. Figure 3(c) collects a wide range of equi-energy plot of HMM cavity resonance structures with inner core radius $h_1$ and outer layer thickness $h_2$. From the dashed line in (c) we got multiple setups all of which support WGM at resonance of 1.79 eV, which is used in figure 3(d) to show $\kappa_{1e}$ and $\kappa_c$ as a function of inner core radius $h_1$. Note that the coupling strength $\kappa_{1e}$ for fixed number of exciton change greatly as the structure mode volume and Purcell factor change, as on the dashed line in figure 3(c), point A: $V_m = 8.99 \times 10^{-5} \left(\frac{\lambda}{n}\right)^3$, $F_P = 4.27 \times 10^4$ and point B: $V_m = 8.62 \times 10^{-5} \left(\frac{\lambda}{n}\right)^3$, $F_P = 4.15 \times 10^4$. However, the critical strength $\kappa_c$ of the coupling stays almost constant, which inspires us that the threshold number of exciton for strong coupling might be constant in these sets of structures. We relate this definition $\kappa_{1e} = \sqrt{N}\mu_J \sqrt{\frac{\hbar\omega}{2\varepsilon\varepsilon_0 V_m}}$ [15, 37] with the TCMT fitting result parameter $\kappa_{1e}$, thus the number of molecule $N$ involved in the threshold strong coupling ($\kappa_{1e} = \kappa_c$) interaction with the hyperbolic metamaterial cavity is,

$$N = \frac{2\varepsilon\varepsilon_0 V_m}{\hbar\omega\mu_J^2}\kappa_c^2$$

Here $\mu_J$ is the molecule dipole moment 20 Debye, $\kappa_c$ is parameter from the Fano fitting from extinction spectrum of the HMM cavity structure without bringing into molecule exciton, $\varepsilon = 1.25^2$ is the relative permittivity of the dielectric layer molecule sit in, $V_m$ is the mode volume, $\hbar\omega = 1.79 eV$ is the resonance frequency. Thanks to the Fano loss in the hyperbolic cavity is associated with the intrinsic permittivity tensor, the molecule exciton number for threshold strong coupling remain 4, after taking rounded number of equation above, which is very close to the final goal of single exciton strong coupling. Therefore, we have discovered a set of tunable hyperbolic metamaterial cavity that requires only 4 molecule exciton to fulfill strong coupling. It is worthwhile to note that future cavity improvement would be possible to further minimize the threshold number of exciton.

## 3. Conclusions

To conclude, we prove that the strong coupling between hyperbolic metamaterial cavity and molecule exciton can benefit from the localized field at the dielectric layer with extremely small mode volume. Strong interaction between cavity mode and molecule exciton are further demonstrated by Rabi splitting in extinction and absorption spectrum, and the anti-crossing phenomenon when the resonance is detuned by varying inner core radius or the outermost layer thickness. Using two-oscillator temporal coupled mode theory, we proved the coupling between two oscillators have reached strong coupling regime. Within multiple setups that display the resonance frequency at 1.79 eV, we are able to extract the threshold number of molecule exciton in each setup by relating the number of molecule exciton with the critical strength. Though the coupling strength in different setups vary a lot with fixed exciton number, we can get the threshold number of exciton for strong coupling with the cavity always around 4, thus realizing few molecule exciton strong coupling with tunable hyperbolic metamaterial structures. Looking forward, experimental designs can fabricate the structures in our designs and facilitate the future study of quantum information.